# MCMC for doubly-intractable distributions

**Iain Murray**
Gatsby Computational
Neuroscience Unit
University College London
i.murray@gatsby.ucl.ac.uk

**Zoubin Ghahramani**
Department of Engineering
University of Cambridge
zoubin@eng.cam.ac.uk

**David J. C. MacKay**
Cavendish Laboratory
University of Cambridge
mackay@mrao.cam.ac.uk

## Abstract

Markov Chain Monte Carlo (MCMC) algorithms are routinely used to draw samples from distributions with intractable normalization constants. However, standard MCMC algorithms do not apply to *doubly-intractable* distributions in which there are additional parameter-dependent normalization terms; for example, the posterior over parameters of an undirected graphical model. An ingenious auxiliary-variable scheme (Møller et al., 2004) offers a solution: exact sampling (Propp and Wilson, 1996) is used to sample from a Metropolis–Hastings proposal for which the acceptance probability is tractable. Unfortunately the acceptance probability of these expensive updates can be low. This paper provides a generalization of Møller et al. (2004) and a new MCMC algorithm, which obtains better acceptance probabilities for the same amount of exact sampling, and removes the need to estimate model parameters before sampling begins.

## 1 Introduction

Markov Chain Monte Carlo (MCMC) methods draw correlated samples from a distribution of interest,

$$p(y|\theta) = f(y;\theta)/\mathcal{Z}(\theta), \tag{1}$$

and use these samples to construct estimators. The normalization $\mathcal{Z}(\theta) = \int f(y;\theta)\,\mathrm{d}y$ is often unknown; standard MCMC methods are designed for use with intractable distributions. The Markov-chain update rule restricts each move to a manageable subset of the $y$'s state space: in Metropolis–Hastings only two settings are considered, the current setting $y$ and a randomly chosen proposal, $y'$; Gibbs sampling changes only one component of $y$ at a time. Metropolis requires an ability to evaluate $f(y;\theta)$ for various $y$, and Gibbs sampling requires the ability to sample from the conditional distributions defined by $f$, but neither method needs to know the normalizing constant $\mathcal{Z}(\theta)$.

We now consider sampling from the posterior over parameters, $\theta$, rather than the variables, $y$. Given a prior $p(\theta)$, the posterior is

$$p(\theta|y) = \left(\frac{f(y;\theta)p(\theta)}{\mathcal{Z}(\theta)}\right)\Big/ p(y). \tag{2}$$

For many models of interest $\mathcal{Z}(\theta)$ cannot be computed. This includes learning in many interesting models, e.g. large-tree-width undirected graphical models and some spatial point processes. Although $p(y)$ is not needed, the normalizing 'constant' $\mathcal{Z}(\theta)$ can not be ignored as it is a function of the parameters.[1] Almost all known valid MCMC algorithms for $\theta$ require an ability to compute values or ratios of $\mathcal{Z}(\theta)$ for the parameter settings considered at each iteration. As MCMC estimators are approximations unless an infinite number of iterations are performed, and each iteration requires an infeasible computation, we call $p(\theta|y)$ a *doubly-intractable* distribution.

In previous work we conjectured that for general undirected models, there are no tractable MCMC schemes giving the correct equilibrium distribution over parameters (Murray and Ghahramani, 2004). Our pragmatic solution was to explore a variety of approximations for the unknown normalizers and their ratios. Such approximations can give useful results, but will not lead to a Markov chain with the correct stationary distribution. This can cause problems in practice.

An ingenious approach by Møller et al. (2004) describes an "efficient Markov chain Monte Carlo method for distributions with intractable normalising constants", which we will refer to as the single aux-iliary variable method (SAVM). Here *efficient* means that in some situations the algorithm will be feasible, unlike all other valid MCMC methods of which we are aware. Our conjecture still stands, but SAVM offers an exciting way out for distributions in which exact sampling is possible. Here, we generalize SAVM and then construct a new method which is easier to apply and has better performance.

[1] This potential source of confusion suggests favoring the statistical physics term *partition function*.

In section 2 we review the SAVM method from Møller et al. (2004). Our interpretation suggests a generalization, described in section 3, which should improve acceptance at a modest cost (section 5). We then introduce a new family of algorithms in section 4, which seem preferable to the SAVM variants.

## 2 The Single Auxiliary Variable Method

In this section we describe the SAVM method due to Møller et al. (2004). We focus our attention on a joint distribution of the form (figure 1(a)):

$$p(y, \theta) = p(y|\theta)p(\theta) = \frac{f(y;\theta)}{\mathcal{Z}(\theta)}p(\theta). \tag{3}$$

Here we assume that $p(\theta)$ is a simple distribution and that $y$ is a vector containing all variables with soft mutual constraints (e.g. the state of an undirected graphical model). We will assume that the $y$ are observed. Note that unobserved $y$, or children of $y$ would not cause any special difficultly: unlike $\theta$ these variables could be sampled with standard MCMC and would otherwise behave like the $y$ that are observed. See Møller et al. (2004) for a more detailed discussion of special cases.

The standard Metropolis–Hastings (M–H) algorithm (Hastings, 1970) constructs a Markov chain through proposals drawn from an arbitrary distribution $q$.

**Metropolis–Hastings Algorithm**

Input: initial setting $\theta$, number of iterations $T$

1. **for** $t = 1 \ldots T$
2. Propose $\theta' \sim q(\theta'; \theta, y)$
3. Compute $a = \frac{p(\theta'|y)q(\theta;\theta',y)}{p(\theta|y)q(\theta';\theta,y)}$
4. Draw $r \sim \mathrm{Uniform}[0, 1]$
5. **if** $(r < a)$ **then** set $\theta = \theta'$.
6. **end for**

Ideal proposals, $q(\theta'; \theta, y)$, would be constructed for rapid exploration of the posterior $p(\theta'|y)$. However, simple perturbations such as a Gaussian with mean $\theta$ are commonly chosen. The accept/reject step, 5 in the algorithm, corrects for the mismatch between the proposal and target distributions.

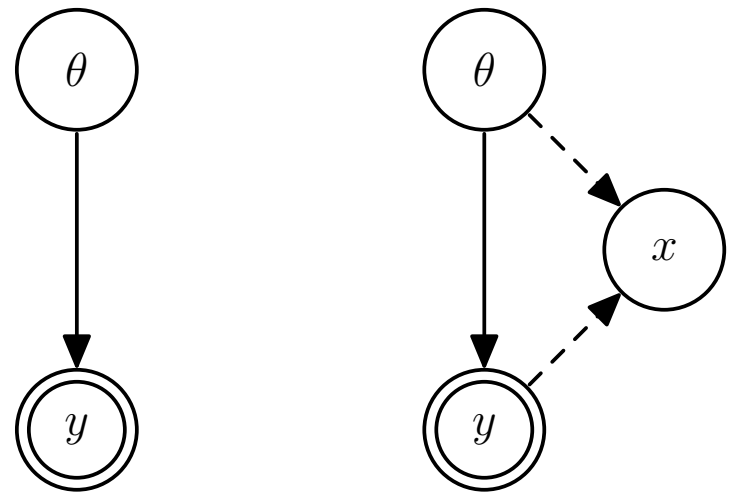

(a) (b)

Figure 1: (a) the original model, unknown parameters $\theta$ generated observed variables $y$, (b) the SAVM augmented model. The conditional distribution of $x$ must have a tractable $\theta$ dependence. In existing approaches this distribution is only a function of one of $y$ or $\theta$, e.g. $f(x; \hat{\theta}(y))/\mathcal{Z}(\hat{\theta})$ or a normalizable function of $(x, \theta)$.

There appears to be no practical way to implement the standard M–H algorithm for doubly-intractable distributions. The acceptance ratio becomes

$$a = \frac{p(\theta'|y)}{p(\theta|y)}\frac{q(\theta;\theta',y)}{q(\theta';\theta,y)} = \frac{f(y;\theta')p(\theta')q(\theta;\theta',y)}{f(y;\theta)p(\theta)q(\theta';\theta,y)} \cdot \frac{\mathcal{Z}(\theta)}{\mathcal{Z}(\theta')}. \tag{4}$$

Perhaps the requirement to compute $\mathcal{Z}(\theta)$ can be removed? We are not free to change $f$, which defines our model. In theory the proposals could be defined to remove explicit $\mathcal{Z}$ dependence from (4), but in practice this does not seem to help: e.g. $q(\theta'; \theta, y) = \mathcal{Z}(\theta)g(\theta')$ or $q(\theta'; \theta, y) \propto 1/\mathcal{Z}(\theta')$ would be difficult to construct without knowing $\mathcal{Z}$, and would be terrible proposals.

Instead Møller et al. (2004) extend the joint distribution to include an auxiliary variable, $x$, which shares the same state space as $y$ (figure 1(b)):

$$p(x, y, \theta) = p(x|\theta, y)\frac{f(y;\theta)}{\mathcal{Z}(\theta)}p(\theta). \tag{5}$$

The joint distribution $p(y, \theta)$ is unaffected. No known method of defining auxiliary variables removes $\mathcal{Z}(\theta)$ from the joint distribution. However, through careful choice of $q$, explicit $\mathcal{Z}(\theta)$ dependence can be removed from the M–H ratio for this distribution:

$$a = \frac{p(x', \theta'|y)q(x, \theta; x', \theta', y)}{p(x, \theta|y)q(x', \theta'; x, \theta, y)}\ . \tag{6}$$

A convenient form of proposal distribution is

$$q(x', \theta'; x, \theta, y) = q(\theta'; \theta, y)q(x'; \theta'), \tag{7}$$

which corresponds to the usual change in parameters $\theta \to \theta'$, followed by a choice for the auxiliary variable. If this choice, which happens to ignore the old $x$, uses

$$q(x'; \theta') = f(x'; \theta')/\mathcal{Z}(\theta')\ , \tag{8}$$

where $f$ and $\mathcal{Z}$ are the same functions as in $p(y|\theta)$, equation (1), then the M–H acceptance ratio becomes

$$a = \frac{p(x'|\theta', y)}{p(x|\theta, y)} \frac{p(\theta'|y)}{p(\theta|y)} \frac{q(x; x', \theta)}{q(x'; x, \theta')} \frac{q(\theta; \theta', y)}{q(\theta'; \theta, y)}$$

$$= \frac{p(x'|\theta', y)}{p(x|\theta, y)} \frac{\mathcal{Z}(\theta) f(y; \theta') p(\theta')}{\mathcal{Z}(\theta') f(y; \theta) p(\theta)} \frac{f(x; \theta) \mathcal{Z}(\theta')}{f(x'; \theta') \mathcal{Z}(\theta)} \frac{q(\theta; \theta', y)}{q(\theta'; \theta, y)}$$

$$= \frac{f(y; \theta') p(\theta')}{f(y; \theta) p(\theta)} \frac{q(\theta; \theta', y)}{q(\theta'; \theta, y)} \cdot \frac{p(x'|\theta', y)}{p(x|\theta, y)} \frac{f(x; \theta)}{f(x'; \theta')} . \quad (9)$$

Now every term can be computed. The big assumption is that we can draw independent, exact samples from the proposal distribution (8). Surprisingly this is possible for some interesting distributions over large numbers of variables with undirected constraints (Propp and Wilson, 1996). The algorithms typically require tracking sets of states through a random, possibly large, number of Markov chain steps; see (Wilson, 1998) for reviews and examples.

The missing part of this description was the conditional distribution of the auxiliary variable $p(x|\theta, y)$. This choice is not key to constructing a valid M–H algorithm but our choice will have a strong impact on the efficiency of the Markov chain. Normally we have a choice over the proposal distribution. Here that choice is forced upon us and instead we choose the target distribution $p(x|y, \theta)$ to match the proposals as closely as possible. We can not maximize the acceptance rate by choosing $p(x|y, \theta) = f(x; \theta)/\mathcal{Z}(\theta)$, as that would reintroduce explicit $\mathcal{Z}(\theta)$ terms into the M–H ratio. Two possibilities are 1) use a normalizable approximation to the ideal case, 2) replace $\theta$ with a fixed value

$$p(x|\theta, y) = p(x|y) = p(x|\hat{\theta}(y)) = \frac{f(x; \hat{\theta})}{\mathcal{Z}(\hat{\theta})} , \quad (10)$$

where $\hat{\theta}$ is a point estimate of the parameters, such as the maximum pseudo-likelihood estimate based on the observations $y$. When the normalization is fixed, it will cancel in (9). The broken lines in figure 1(b) indicate that while $x$ could be a child of $\theta$ and $y$, in practice all previous methods have used only one of the possible parents. For concreteness we assume $p(x|\theta, y) = f(x|\hat{\theta})/\mathcal{Z}(\hat{\theta})$ for some fixed $\hat{\theta}(y)$ in all that follows, but our results are applicable to either case.

## 3 A tempered-transitions refinement

The M–H rule using the acceptance rate (9) must implicitly estimate the relative importance of pairs of states; in some sense it is using the additional random variables to approximate the acceptance ratio (4). This becomes apparent by identifying two unbiased one-sample importance-sampling estimators:

$$\frac{\mathcal{Z}(\hat{\theta})}{\mathcal{Z}(\theta')} \approx \frac{f(x'; \hat{\theta})}{f(x'; \theta')} \qquad x' \sim f(x; \theta')/\mathcal{Z}(\theta') \quad (11)$$

$$\frac{\mathcal{Z}(\hat{\theta})}{\mathcal{Z}(\theta)} \approx \frac{f(x; \hat{\theta})}{f(x; \theta)} \qquad x \sim f(x; \theta)/\mathcal{Z}(\theta) \quad (12)$$

A biased estimate of $\mathcal{Z}(\theta)/\mathcal{Z}(\theta')$ is obtained by dividing (11) by (12). The unknown constant $\mathcal{Z}(\hat{\theta})$ fortuitously cancels. Unlike previous attempts, substituting this elementary approximation into the M–H rule (4) leads to a valid algorithm.

Given the simple nature of SAVM's "importance sampling" estimators, or equivalently the mismatch between $p(x|\theta, y)$ and $q(x; \theta, y)$, the M–H algorithm can suffer a high rejection rate. Annealed importance sampling (AIS) (Neal, 2001; Jarzynski, 1997) is a natural way to make the target and proposal distributions closer to improve estimators of normalizing constants. Linked importance sampling (Neal, 2005) could also be used as a drop-in replacement. We now show that this is a valid extension of SAVM.

We notionally extend the auxiliary variables $x$ to an ensemble of similar variables $X = \{x_1, x_2, ...x_{K+1}\}$ (figure 2). We give $x_1$ the same conditional distribution (10) as the single auxiliary variable $x$ in SAVM. The distribution over the remaining variables is defined by a sequence of Markov chain transition operators $\tilde{T}_k(x_{k+1}; x_k)$ with $k = 1 \ldots K$:

$$\begin{aligned} p(x_2|x_1, \theta, y) &\sim \tilde{T}_1(x_2; x_1, \theta, \hat{\theta}(y)) \\ p(x_3|x_2, \theta, y) &\sim \tilde{T}_2(x_3; x_2, \theta, \hat{\theta}(y)) \\ &\cdots \\ p(x_{K+1}|x_K, \theta, y) &\sim \tilde{T}_K(x_{K+1}; x_K, \theta, \hat{\theta}(y)). \end{aligned} \quad (13)$$

Transition operator $\tilde{T}_k$ is chosen to leave a distribution $p_k$ stationary. The sequence of stationary distributions should bridge from the approximate or estimator-based distribution $p(x_1|\theta, y)$ towards the distribution $f(x; \theta)/\mathcal{Z}(\theta)$ from which are forced to draw an exact sample as part of the proposal. One interpolation scheme is:

$$p_k(x; \theta, \hat{\theta}) \propto f(x; \hat{\theta})^{\beta_k} f(x; \theta)^{(1-\beta_k)} \equiv f_k(x; \theta, \hat{\theta}). \quad (14)$$

The sequence $$\beta_k = \frac{K - k + 1}{K + 1} \quad (15)$$

is used in Neal (2004) as a default choice. Note that as with AIS, the final stationary distribution, $p_K$, is nearly the same as $f(x; \theta)/Z(\theta)$ for large $K$. Other sets of bridging distributions may perform better, although finding them may be difficult (Meng and Wong, 1996).

We now perform M–H sampling on the new joint distribution $p(\theta, X|y)$. First we propose a change in parameters, $\theta' \sim q(\theta'; \theta, y)$, as before. Then a change in $X$ is proposed in "reverse order": $x_{K+1}$ is drawn by exact sampling from the same distribution as in SAVM and $x_K \ldots x_1$ are proposed using transition operators that bridge towards $p(x_1|\theta', y)$:

$$\begin{aligned} q(x_{K+1}; \theta', y) &= f(x_{K+1}; \theta')/\mathcal{Z}(\theta') \\ &\equiv p_{K+1}(x_{K+1}|\theta', \hat{\theta}(y)) \\ q(x_K; x_{K+1}, \theta', y) &\sim T_K(x_K; x_{K+1}, \theta', \hat{\theta}(y)) \\ q(x_{K-1}; x_K, \theta', y) &\sim T_{K-1}(x_{K-1}; x_K, \theta', \hat{\theta}(y)) \\ &\cdots \\ q(x_1; x_2, \theta', y) &\sim T_1(x_1; x_2, \theta', \hat{\theta}(y)) \ , \end{aligned} \tag{16}$$

where $T_k$ are the corresponding reverse transition operators to those used to define $p(X|\theta, y)$, such that

$$T_k(x'; x)p_k(x) = \tilde{T}_k(x; x')p_k(x') \ . \tag{17}$$

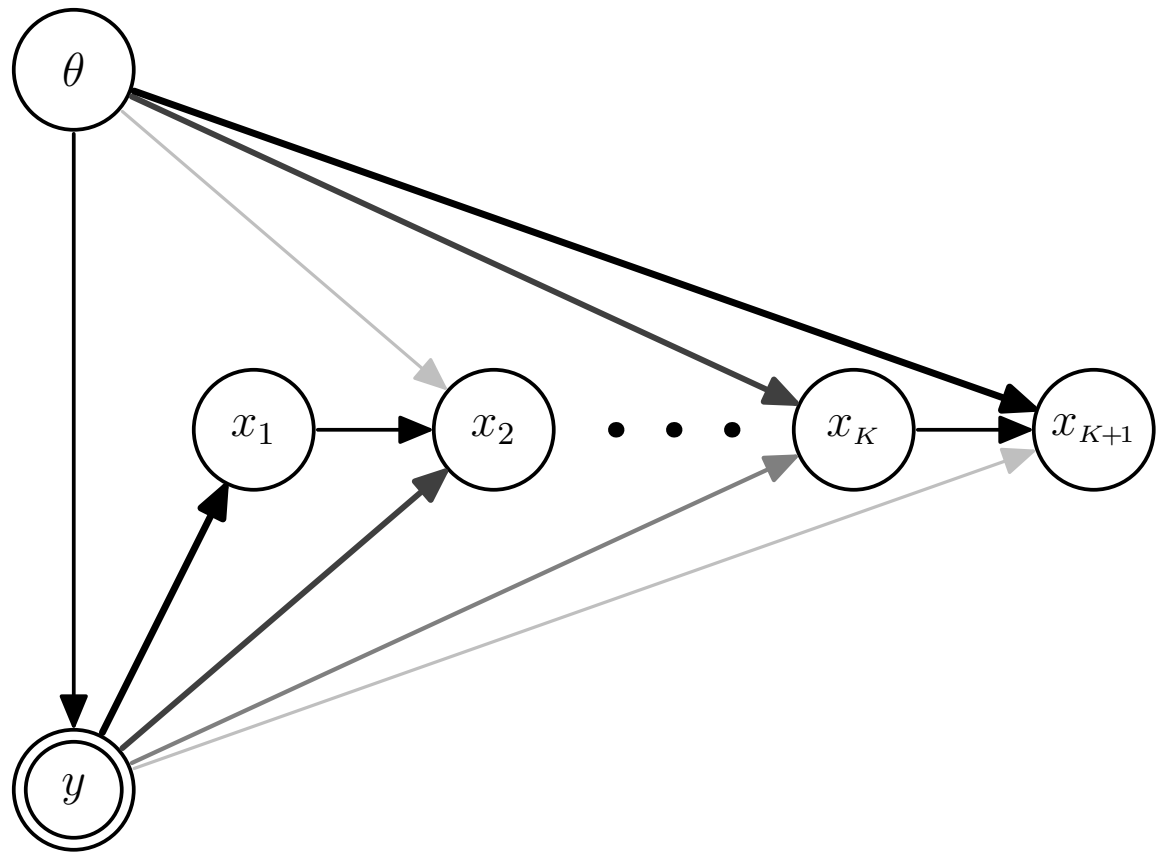

Figure 2: The joint distribution for the annealing-based multiple auxiliary variable method (MAVN). Here it is assumed that $p(x_1|\theta, y)$ is based only on a data-driven parameter estimate as in (10). The auxiliary variables bridge towards the distribution implied by $\theta$. The gray-level and thickness of the arrows from $y$ and $\theta$ indicate the strengths of influence, $\beta_k$, on the auxiliary variables in (14).

The M–H ratio for accepting the whole move $(\theta, X = \{x_1, x_2, ...\}) \to (\theta', X' = \{x'_1, x'_2, ...\})$ is still free of any explicit $\mathcal{Z}(\theta)$-dependence. Substituting equations (13) and (16) into (6), eliminating $\tilde{T}$ with (17), rearranging and cancelling gives:

$$\begin{aligned} a = \frac{f(y; \theta')p(\theta')}{f(y; \theta)p(\theta)} \frac{q(\theta; \theta', y)}{q(\theta'; \theta, y)} \cdot \\ \prod_{k=0}^{K} \frac{f_k(x'_{k+1}; \theta', \hat{\theta})}{f_{k+1}(x'_{k+1}; \theta', \hat{\theta})} \frac{f_{k+1}(x_{k+1}; \theta, \hat{\theta})}{f_k(x_{k+1}; \theta, \hat{\theta})} \ . \end{aligned} \tag{18}$$

The terms have been arranged to make it clear that all ratios involving the auxiliary variables can be computed online as the transitions $T$ are generated. As in SAVM ($K = 0$), all unknown normalization constants cancel. We call this method with $K \geq 1$ the multiple auxiliary variable method (MAVM).

While we were motivated by improving the importance sampling like estimators using AIS, it turns out MAVM is more closely related to "Tempered Transitions" (Neal, 1996). Our approach has cheaper moves than standard tempered transitions, which would regenerate $x_1 \ldots x_{K+1}$ from $p(X|\theta, y)$ before every M–H proposal. This trick could be applied to tempered transitions in general; the trade-off with acceptance rate will be explored in future work.

In reviewing SAVM it appeared that the auxiliary proposal distribution had to consist of a single exact sample. Our extension allows us to augment the sample with a fixed number of additional steps. This allows us to improve the implicit normalization constant estimation and improve the acceptance rate, for some additional cost. However, no further expensive exact sampling, on top of that needed by the original algorithm, is required per iteration. The performance as a function of $K$ is explored in section 5.

We have also provided an answer to an open question in Møller et al. (2004) on how to use both $\theta$ and $y$ in $p(x|\theta, y)$. We use $y$ in coming up with the point estimate of the parameters to get the distribution in roughly the right place. Then we bridge towards a better fit for $\theta$ using tempered transitions.

## 4 Simpler, more direct methods

We know the SAVM algorithm is valid, as it is an implementation of M–H. And we have a pseudo-explanation in terms of importance sampling, which motivated MAVM. However, the meaning of the auxiliary variables has been left unexplained. It is tempting to draw parallels with alternative algorithms, e.g. the Boltzmann machine learning rule (Ackley et al., 1985) also involves generating variables that exist in the same state space as the observed variables. However, it seems difficult to attach any general meaning to the auxiliary variable settings visited by the Markov chain. The correlated samples come asymptotically from $\int p(x|y, \theta)p(\theta)\, \mathrm{d}\theta$, which can be arbitrary. In this section we derive a method which draws more meaningful variables. In section 5 we will see that our simpler method leads to an improvement in performance.

It was unclear why two normalizing constant ratio estimates were needed as a proxy for $\mathcal{Z}(\theta)/\mathcal{Z}(\theta')$. A more direct estimate is obtained from a single one-sample unbiased importance sampler:

$$\frac{\mathcal{Z}(\theta)}{\mathcal{Z}(\theta')} \approx \frac{f(x; \theta)}{f(x; \theta')} \qquad\qquad x \sim f(x; \theta')/\mathcal{Z}(\theta'). \tag{19}$$

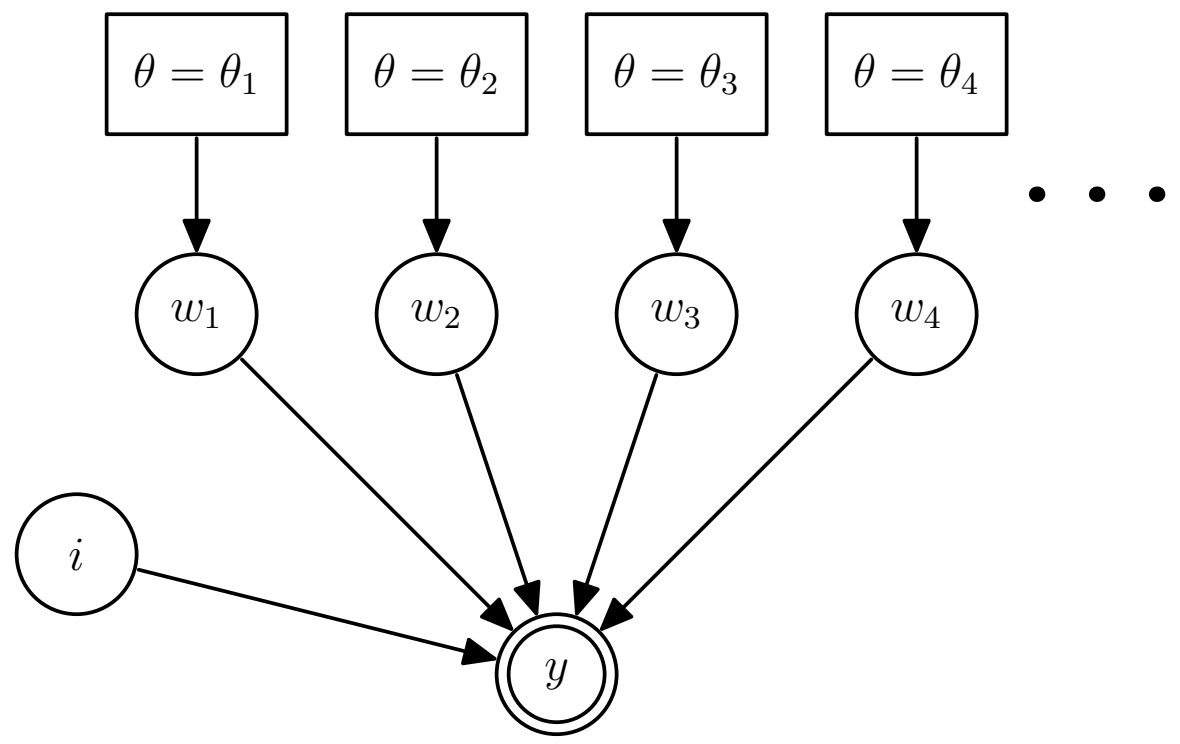

Figure 3: An alternative representation of the generative model for observations $y$. All possible parameter settings, $\theta_l$, are instantiated, fixed and used to generate a set of data variables $w_l$. The indicator $i$ is used to set $y = w_i$. The posterior over $\theta_i$, the parameter chosen by the indicator variable $i$, is identical to $p(\theta|y)$ in the original model.

In this section we provide a proof that using this direct estimator leads to a valid algorithm. The work in this section was originally inspired by relating Carlin and Chib (1995) to SAVM. The joint model we use is also related to parallel or replica tempering methods in the physics literature, e.g. Swendsen and Wang (1986).

Consider a huge model in which all possible parameter settings $\theta_l$ exist. Each parameter setting generates its own setting of variables

$$w_l \sim f(w_l;\theta_l)/\mathcal{Z}(\theta_l). \qquad (20)$$

To generate the data, $y$, we choose an index using the same prior over parameters as in our original model $i \sim p(i) = p(\theta_i)$ , and copy $y$ from $w_i$: $p(y) = \delta(y - w_i)$.

This generative model is illustrated in figure 3. The marginal distribution over $y$ is identical to that of the original model. Also the prior and posterior over which parameter $\theta_i$ generated the data is equivalent to the distributions over the original $\theta$. All that is different is that we also choose to generate a lot of unobserved data, $\{w_{l\neq i}\}$, which does not affect the marginal distribution over the variables of interest.

If the parameters take on a finite number of possible settings, standard MCMC would now apply. All normalizing constants, $\mathcal{Z}_l(\theta_l)$, appear in the joint distribution for all choices of the index $i$ and therefore cancel in M–H ratios. However, all the variables $w_l$ must be instantiated and sampled for sufficiently long to reach equilibrium. Naïvely it seems this is a very costly or impossible approach. However, we now outline how, through judicious use of exact sampling, it is possible to deal with this model even in the limit of an infinite number of possible parameter settings.

Consider starting a chain with index $i$ and proposing a change to index $j$ with probability $q(j;i)$. In order to ensure that $w_j = y$, we deterministically swap the settings of $w_i$ and $w_j$. The new and old states of the joint model have probability ratio:

$$\frac{p(\{w'_l\}, j)}{p(\{w_l\}, i)} = \frac{p(\theta_j) f(y;\theta_j) f(w_j;\theta_i) \cancel{\prod_{l\neq i,j} f(w_l;\theta_l)}}{p(\theta_i) f(y;\theta_i) f(w_j;\theta_j) \cancel{\prod_{l\neq i,j} f(w_l;\theta_l)}}\,.$$

As all terms involving $w_{l\neq i,j}$ cancel, we need only know the initial setting of $w_j$. Under the joint model $p(w_j) = f(w_j;\theta_j)/\mathcal{Z}(\theta_j)$; we can generate $w_j$ when required by exact sampling. This is not part of the proposal, which was deterministic after the change of index $i \to j$. We simply deferred finding out what $w_j$ was until it was needed. The M–H ratio becomes

$$a = \frac{q(i;j)p(\theta_j)f(y;\theta_j)}{q(j;i)p(\theta_i)f(y;\theta_i)} \cdot \frac{f(w_j;\theta_i)}{f(w_j;\theta_j)} \qquad (21)$$

for which all terms are tractable. Loosely speaking the second term acts as the importance sampling estimate suggested at the beginning of the section. Compare this to the last term of (9) for SAVM.

Despite the underlying infinite model, the resulting algorithm is slightly simpler than the original SAVM:

**Single-variable Exchange algorithm**

Input: initial $\theta$, number of iterations $T$

**for** $t = 1 \ldots T$
  Propose $\theta' \sim q(\theta';\theta, y)$
  generate an auxiliary variable,
    $w \sim f(w;\theta')/\mathcal{Z}(\theta')$
  Compute
$$a = \frac{q(\theta;\theta', y)p(\theta')f(y;\theta')}{q(\theta';\theta, y)p(\theta)f(y;\theta)} \cdot \frac{f(w;\theta)}{f(w;\theta')} \qquad (22)$$
  Draw $r \sim \text{Uniform}[0,1]$
  **if** $(r < a)$ **then** set $\theta = \theta'$.
**end for**

We call this the exchange algorithm. Each step tries to take the data $y$ from the current parameter setting $\theta$. We speculate that a better parameter setting is $\theta'$, which was generated by $q(\theta';\theta, y)$. How can we persuade $\theta$ to give up the data to the rival parameter setting $\theta'$? We offer it a replacement data set $w$ from $\theta'$s distribution. If $f(w;\theta)/f(y;\theta) > 1$ then this replacement is preferred by $\theta$ to the real data $y$, which is a good thing. We have to consider both sides of the exchange: the ratio $f(y;\theta')/f(w;\theta')$ measures how much $\theta'$ likes the trade in data sets. The other terms in (22) respect any disparity between the frequency with which we propose swaps and our prior preference over the parameter that should own the data.

The exchange algorithm is a valid MCMC method because it is the M–H algorithm for a joint system with the correct posterior distribution $p(\theta_i|y)$. We now outline a more direct mathematical proof; see Neal (2004,

p3) for the details of a similar proof. It is easily shown that detailed balance holds for any particular intermediate exact sample $w$ by checking that the probability of starting at $\theta_i$ (under the intended equilibrium distribution $p(\theta_i|y)$) and then moving to $\theta_j$ via the exact sample $w$ is symmetric in $i$ and $j$. Summing over the intermediate quantity $w$ gives detailed balance overall.

### 4.1 Reinterpreting SAVM

Seen in the light of the joint distribution in figure 3, the SAVM method appears slightly strange. The SAVM method can be reproduced by augmenting the joint model in figure 3 with the SAVM auxiliary variable, $x$, and using the following proposal:

1. Draw $j \sim q(j;i)$
2. Deterministically perform the three-way swap $x = w_j$, $w_i = x$ and $w_j = y$.

The acceptance factor for this proposal is precisely the M–H ratio in (9). If we want to take $y$ from $\theta_i$ and give it to rival setting $\theta_j$ why involve a third parameter $\hat{\theta}$? In section 5 we see that the third party can make the transaction harder (figure 6) or mediate it (figure 4). In the next section we add auxiliary variables to the exchange algorithm that are specifically designed to make the swap more palatable.

### 4.2 Bridging

In section 3 we improved SAVM's proposal distribution by bridging between the original proposal and target distributions. A similar refinement can be applied to the exchange algorithm. "After taking the new parameter's data we take steps to make it more appealing to the current parameter." The general exchange algorithm with bridging is shown opposite; $K=0$ recovers the previous single-variable version.

This bridging method is slightly less general than our extension to SAVM, as the transition operators $R$ must satisfy detailed balance:

$$\begin{aligned} &R_k(x';x,\theta,\theta')p_k(x;\theta,\theta') \\ &\quad = R_k(x;x',\theta,\theta')p_k(x';\theta,\theta'), \end{aligned} \tag{25}$$

where the stationary distributions $p_k$ and corresponding $f_k$ are defined as before in equations (14) and (15), i.e. $p_k(x;\theta,\theta') \propto f_k(x;\theta,\theta') = f(x;\theta')^{\beta_k} f(x;\theta)^{1-\beta_k}$.

For clarity the details of the motivating infinite model, section 4, have been omitted from the algorithm. The correspondence is as follows: the exact sample $x_0 \sim p_0(x_0;\theta,\theta')$ is the current setting of $w'$ corresponding to $\theta'$, which we then make more attractive to $\theta$ through a sequence of transitions before proposing to set $w = x_K$ and $w' = y$.

**Exchange algorithm with bridging**

Input: initial $\theta$, #iterations $T$, #bridging levels $K$

1. **for** $t = 1 \ldots T$
2. Propose $\theta' \sim q(\theta';\theta,y)$
3. generate $K+1$ auxiliary variables:

$$\begin{aligned} x_0 &\sim p_0(x_0;\theta,\theta') \\ &\equiv f(x_0;\theta')/\mathcal{Z}(\theta') \\ x_1 &\sim R_1(x_1;x_0,\theta,\theta') \\ x_2 &\sim R_2(x_2;x_1,\theta,\theta') \\ &\cdots \\ x_K &\sim R_K(x_K;x_{K-1},\theta,\theta') \end{aligned} \tag{23}$$

4. Compute

$$a = \frac{q(\theta;\theta',y)p(\theta')f(y;\theta')}{q(\theta';\theta,y)p(\theta)f(y;\theta)} \cdot \prod_{k=0}^{K} \frac{f_{k+1}(x_k;\theta,\theta')}{f_k(x_k;\theta,\theta')} \tag{24}$$

5. Draw $r \sim \text{Uniform}[0,1]$
6. **if** $(r < a)$ **then** set $\theta = \theta'$.
7. **end for**

The proof of validity is again a strong detailed balance condition. Sketch: the probability of being in state $\theta_i$ at equilibrium, obtaining the numbers $(x_0, x_1, x_2, \ldots, x_K)$ and transitioning to state $\theta_j$ is the same as the probability of being in state $\theta_j$ at equilibrium, obtaining the numbers $(x_K, \ldots, x_2, x_1, x_0)$ and transitioning to state $\theta_i$. Summing over all possible settings of intermediate auxiliary variables gives detailed balance overall.

## 5 Comparison of the algorithms

We consider a concrete example for which all computations are easy. This allows comparison with exact partition function evaluation as in (4) and averaging over chains starting from the true posterior. We consider sampling from the posterior of a single precision parameter $\theta$, which has likelihood corresponding to $N$ i.i.d. zero-mean Gaussian observations $y = \{y_1, y_2, \ldots y_N\}$, with a conjugate prior:

$$p(y_n|\theta) = \mathcal{N}(0, 1/\theta), \quad p(\theta|\alpha,\beta) = \text{Gamma}(\alpha,\beta). \tag{26}$$

The corresponding posterior is tractable

$$p(\theta|y) = \text{Gamma}\left(N/2+\alpha,\ \textstyle\sum_n y_n^2/2+\beta\right), \tag{27}$$

but we pretend that the normalizing constant in the likelihood is unknown. We compare the average acceptance rate of the algorithms for two choices of proposal distribution $q(\theta';\theta,y)$.

All of the algorithms require $N$ exact Gaussian samples, for which we used standard generators. We also draw directly from the stationary distributions, $p_k$, in the bridging algorithms. This simulates an ideal case where the energy levels are close, or the transition operators mix well. More levels would be required for the same performance with less efficient operators. We now report results for $\alpha = 1$, $\beta = 1$, $N = 1$ and $y = 1$.

The first experiment uses proposals drawn directly from the parameter posterior (27). The M–H acceptance probability in (4) becomes $a \equiv 1$; all proposals would be accepted if $\mathcal{Z}(\theta)$ were computed exactly. Therefore any rejections are undesirable by-products of the auxiliary variable scheme, which can (implicitly) obtain only noisy estimates of the normalizing constants. Figure 4 shows that both MAVM and the exchange algorithm improve over the SAVM baseline. It appears that a large number, $K$, of bridging levels are required to bring the acceptance rate close to the attainable $a = 1$. However, significant benefit is obtained from a relatively small number of levels, after which there are diminishing returns. As each algorithm requires an exact sample, which in applications can require many Markov chain steps, the improvement from a few extra steps ($K > 0$) can be worth the cost (see section 5.1).

In this artificial situation the performance of MAVM was similar to the exchange algorithm. This result favors the exchange algorithm, which has a slightly simpler update rule and does not need to find a maximum (pseudo)-likelihood estimate before sampling begins. In figure 4 we had set $\hat{\theta} = 1$. Figure 5 shows that the performance of MAVM falls off when this estimate is of poor quality. For moderate $K$, the exchange algorithm automatically obtains an acceptance rate similar to the best possible performance of MAVM; only for $K = 0$ was performance considerably worse than SAVM. For this simple posterior $\hat{\theta}$ sometimes manages to be a useful intermediary, but by $K = 1$ the exchange algorithm has caught up with MAVM.

More importantly, the exchange algorithm performs significantly better than SAVM and MAVM in a more realistic situation where the parameter proposal $q(\theta'; \theta, y)$ is not ideal. Figure 6 shows results using a Gaussian proposal centred on the current parameter value. The exchange algorithm exploits the local nature of the proposal, rapidly obtaining the same acceptance rate as exactly evaluating $\mathcal{Z}(\theta)$. MAVM performs much worse, although adding bridging levels does rapidly improve performance over the original SAVM algorithm. SAVM is now hindered by $\hat{\theta}$, which is more rarely between $\theta$ and $\theta'$.

The posterior distribution over $\theta$, equation (27), be-comes sharper for $N > 1$. This makes the performance of SAVM and MAVM fall off more rapidly as $\hat{\theta}$ is moved away from its optimum value. These methods require better estimates of $\theta$ with larger datasets.

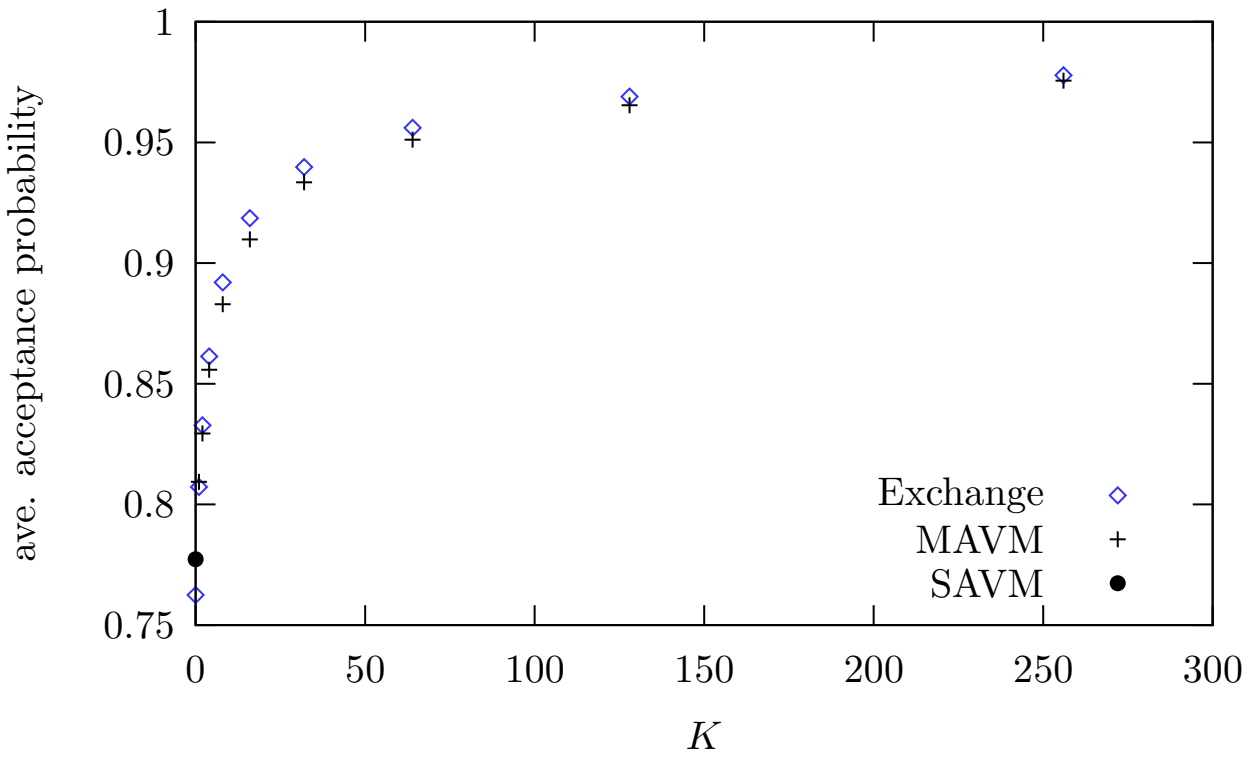

Figure 4: Average acceptance rate as a function of $K$ for the Gaussian example (section 5). MAVM with $K = 0$ corresponds to SAVM, the method of Møller et al. (2004). Exact normalizing constant evaluation in (4) would give an acceptance rate of one.

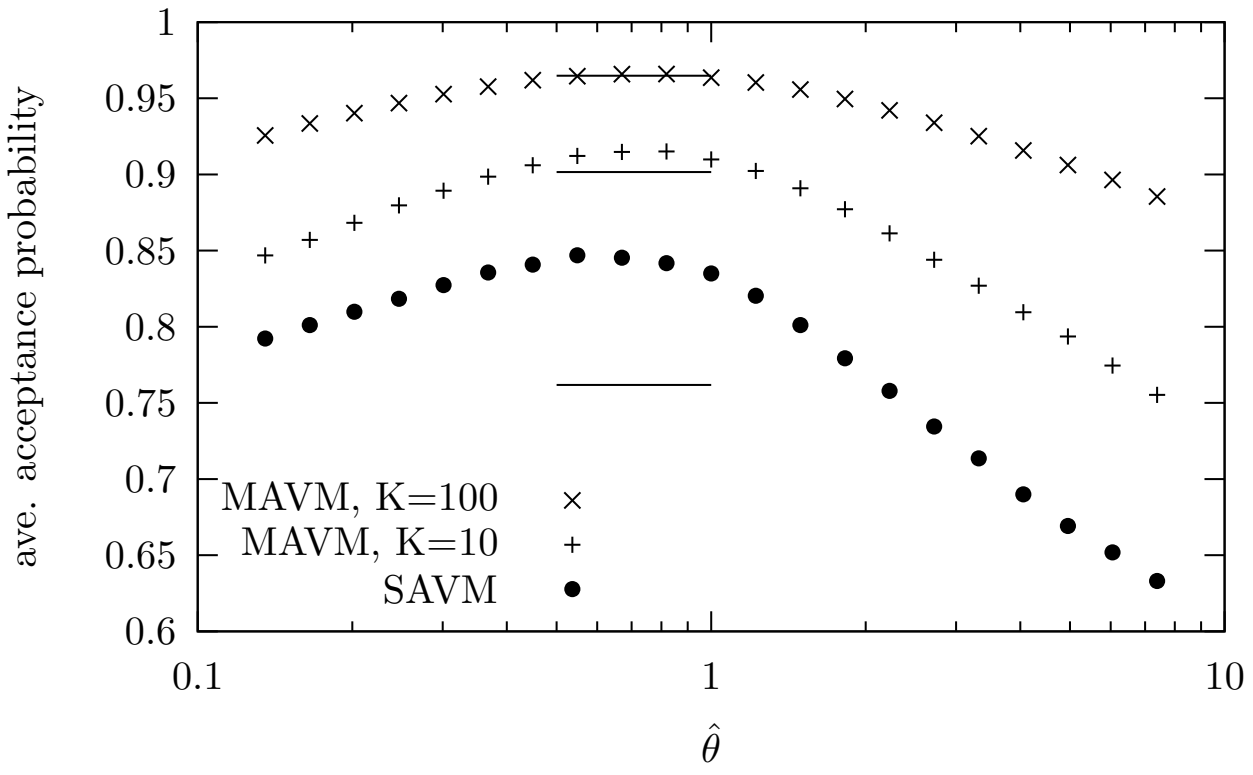

Figure 5: Average acceptance rate under the example in section 5 as a function of the initial parameter estimate required by SAVM ($K = 0$) and our extended version, MAVM. Horizontal bars show the results for the exchange algorithm, which has no $\hat{\theta}$, for $K = 0, 10, 100$.

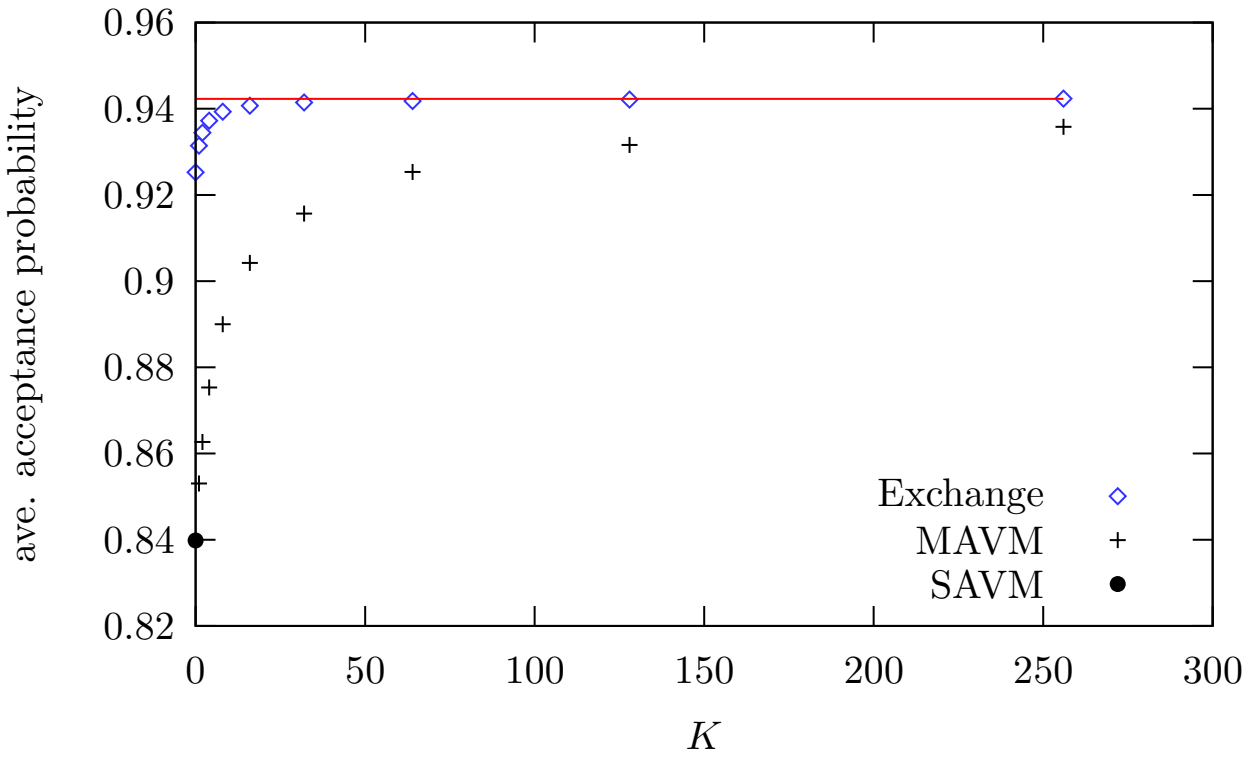

Figure 6: As in figure 4 but with a Gaussian proposal distribution of width 0.1 centered on the current parameter setting. The horizontal line shows the maximum average acceptance rate for a reversible transition operator, this is obtained by exact normalizing constant evaluation in (4).

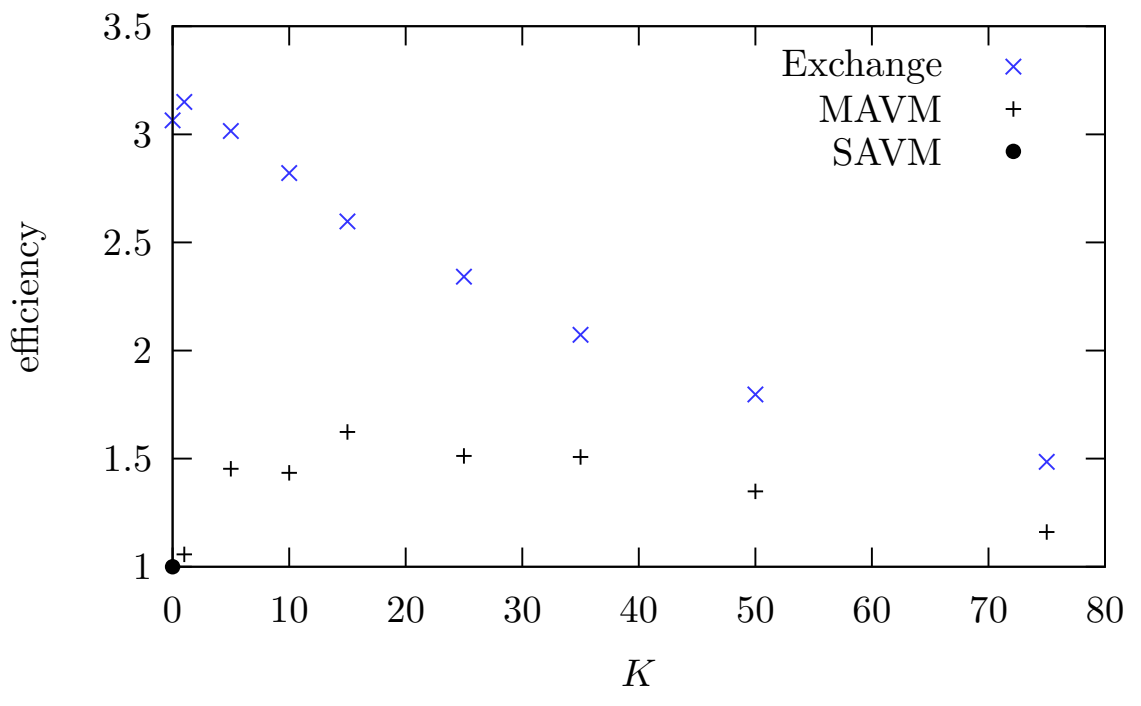

Figure 7: Performance on a $10 \times 30$ toroidal square Ising lattice. The data were generated from an exact sample with $\theta_J = 0.3$ and $\theta_h = 0$. Proposals were Gaussian pertubations of width 0.01. The plot shows efficiency: effective number of samples, estimated by R-CODA (Cowles et al., 2006), divided by total number of Gibbs updates (computer time) normalized to give SAVM an efficiency of one. In this example the exchange algorithm provides better mixing times than SAVM/MAVM for all $K$; bridging improves over SAVM, but is only worth the cost with the exchange algorithm for $K = 1$.

### 5.1 Ising Model

We have also considered the Ising model distribution with $y_i \in \{\pm 1\}$ on a graph with nodes $i$ and edges $E$:

$$p(y|\theta) = \frac{1}{\mathcal{Z}(\theta)} \exp\Big( \sum_{i \neq j \in E} \theta_J y_i y_j + \sum_i \theta_h y_i \Big). \quad (28)$$

As in Møller et al. (2004) we used uniform priors over $|\theta_h| < 1$ and $|0 < \theta_J < 1|$. We used the *Summary States* algorithm (Childs et al., 2001) an implementation of *Coupling from the Past* (Propp and Wilson, 1996) for exact sampling and a single sweep of Gibbs sampling for transition operators $T$ and $R$. Figure 7 suggests that the advances over SAVM introduced in this paper can carry over to more challenging distributions with realistic MCMC transition operators.

## 6 Discussion

MCMC methods typically navigate complicated probability distributions by local diffusion — longer range proposals will be rejected unless carefully constructed. It is usually the case that as the step-size of proposals are made sufficiently small the acceptance rate of a M–H method tends to one. However, SAVM does not have this property, it introduces rejections even when $\theta' = \theta$. While the exchange algorithm has $a \to 1$ for all $K$ as the step-size tends to zero, MAVM will only recover $a = 1$ as $K \to \infty$. This is because the third party in the proposed swap (see section 4.1) is not necessarily close to $\theta$. Even in a simple unimodal 1-dimensional posterior distribution, figure 6, this is a significant disadvantage in comparison with the exchange algorithm. We found the exchange algorithm performs better than the only other existing MCMC method for this problem and is simpler to implement.

**Acknowledgements:** We thank Radford Neal for useful comments.